\documentclass[10pt,a4paper]{article}
\pdfoutput=1

\usepackage[utf8]{inputenc}
\usepackage{uniinput}
\usepackage{jheppubmod}
\usepackage{amsmath, amssymb, tensor}
\usepackage{url}
\usepackage{bm}
\usepackage{mathtools}
\usepackage{relsize}
\usepackage{cancel}
\usepackage{lipsum,cite,mathrsfs}
\usepackage{parskip}
\usepackage[dvipsnames]{xcolor}
\usepackage{soul}
\usepackage{calrsfs}
\usepackage{enumitem}
\usepackage[stable]{footmisc} 

\renewcommand{\d}{\mathrm{d}}

\newcommand{\bea}{\begin{eqnarray}}
\newcommand{\eea}{\end{eqnarray}}
\newcommand{\be}{\begin{equation}}
\newcommand{\ee}{\end{equation}}

\usepackage{microtype}
\allowdisplaybreaks

\title{Shedding light on dark bubble cosmology}
\author[a]{Ivano Basile,}
\emailAdd{ivano.basile@lmu.de}
\affiliation[a]{Arnold Sommerfeld Center for Theoretical Physics, Ludwig Maximilians Universität München,\\ Theresienstrasse 37, 80333 München, Germany.}

\author[b]{Ulf Danielsson,}
\emailAdd{ulf.danielsson@physics.uu.se}
\affiliation[b]{Institutionen för fysik och astronomi,
	Uppsala Universitet, Box 803, SE-751 08 Uppsala, Sweden}

\author[b,c,d]{Suvendu Giri,}
\emailAdd{suvendu.giri@princeton.edu}
\affiliation[c]{Department of Physics, Princeton University, Princeton, New Jersey 08544, USA}
\affiliation[d]{Princeton Gravity Initiative, Princeton University, Princeton, New Jersey 08544, USA}

\author[b]{Daniel Panizo}
\emailAdd{daniel.panizo@physics.uu.se}

\abstract{
Dark bubble cosmology is an alternative paradigm to compactification, which can circumvent issues of moduli stabilization and scale separation. In this paper we investigate how electromagnetic fields can be incorporated in this framework. Worldvolume fields backreact on the ambient universe in which the bubble expands, which in turn affects the energy-momentum distribution and the effective gravity induced on the brane. We compute these effects, showing that the resulting four-dimensional cosmology consistently includes electromagnetic waves.}

\preprint{UUITP-30/23}

\begin{document}

\maketitle

\newpage

\section{Introduction}\label{sec:intro}

Building a realistic model of cosmology and particle physics from string theory is the holy grail of string phenomenology. Many efforts in this direction resulted in great progress over the last few decades, but some key obstacles remain. Three of the most prominent issues are stabilizing moduli, obtaining scale separation for the extra dimensions (once stabilized), and producing a positive and small dark energy. The last of which requires breaking supersymmetry, typically generating instabilities which can be difficult to control.\footnote{See \cite{Danielsson:2018ztv,Cicoli:2018kdo,Berglund:2022qsb} for a review of the difficulties involved in constructing stable dS vacua in string theory.} Dark bubble cosmology \cite{Banerjee:2018qey} constitutes a compelling alternative framework in which these issues may be addressed.\footnote{See \cite{Banerjee:2021yrb, Banerjee:2022myh} for a review of dark bubble cosmology, \cite{Kaloper:1999sm, Shiromizu:1999wj, Vollick:1999uz, Gubser:1999vj} for earlier work discussing cosmology on a 3-brane and brane gravity localization, and \cite{Erickson:2021psj, Leung:2022nhy} for recent work in the context of supergravity.} To begin with, its very existence requires breaking supersymmetry and producing a positive dark energy,\footnote{Other recent approaches towards constructing dS vacua in string theory/M-theory include breaking supersymmetry using negative energy sources like O-planes \cite{Cordova:2018dbb,Cordova:2019cvf} or Casimir energy \cite{DeLuca:2021pej}. See \cite{Berglund:2021xlm} and references within, for a construction closely related to the dark bubble scenario, and \cite{Bandos:2023yyo} for recent work studying a 3-brane mediating the decay between 4D vacua} since its underlying mechanism rests on the decay of a metastable anti-de Sitter (AdS) universe via bubble nucleation.\footnote{See \cite{Koga:2019yzj,Koga:2020jok,Koga:2022opd} for catalytic effects on bubble nucleariton in the presence of black holes or a cloud of strings.} Moreover, scale separation on the bubble universe can be implemented not via small extra dimensions, but rather via quasi-extremal tension-to-charge ratios of the bubble. This scenario reproduces important features of gravitational interactions at long distances \cite{Banerjee:2019fzz, Banerjee:2020wix, Banerjee:2020wov, Banerjee:2021qei, Danielsson:2022fhd}, and it can be constructed in non-supersymmetric settings in string theory \cite{Basile:2020mpt, Danielsson:2022lsl}.

One of the most important problems in dark bubble cosmology, first introduced in \cite{Banerjee:2018qey}, is to understand the uplift of four dimensional matter fields compatible with the standard model of particle physics into the higher dimensions. In \cite{Banerjee:2018qey} it was argued that massless radiation is induced in 4D by heating up the system with the help of a bulk black hole at the centre of the bubble. Furthermore, it was shown that fundamental semi-infinite strings, pulling on the brane from above, give rise to particles with Planckian masses in 4D. Neither of these are promising candidates to represent known matter. As discussed in \cite{Basile:2020mpt,Basile:2021vxh}, to obtain masses compatible with the standard model of particle physics, we would need fundamental strings stretching between branes separated by distances much shorter than the AdS-scale.\footnote{In \cite{Banerjee:2019fzz}, it was speculated how semi-infinite strings with a lower tension could play a similar role.} Given the stringy embedding presented in \cite{Danielsson:2022lsl}, a natural possibility would be to separate the branes on the five internal compact dimensions at a distance much smaller than the size of the dimensions.\footnote{Note that fundamental strings stretching between opposite poles of the sphere in internal dimensions would yield masses of order Planck scale.} 

In this paper we will not consider the massive, or non-Abelian sector of the standard model of particle physics, but instead focus on pure electromagnetism. This is already highly non-trivial, requiring an intricate interplay between the 4D physics induced on the brane and what happens in the bulk. As we will see, the 4D electromagnetic waves will not only involve a gauge field trapped on the bubble, but also crucially excite oscillating Kalb-Ramond fields in the bulk. Together, they make sure that all equations of motion, as well as the 4D and 5D Einstein equations together with Israel's junction conditions across the bubble wall, are solved.

There are interesting similarities, but also differences, compared to the uplift of gravitational waves that was done in \cite{Danielsson:2022fhd}. We therefore start out with a review of these results in section \ref{sec: db_review}. In section \ref{sec: 4d_em_cosmo} we discuss our expectations concerning electromagnetic waves in 4D. Section \ref{sec: uplift} is where we perform the uplift of 4D electromagnetic waves to 5D, and subsequently check that they induce the expected 4D energy momentum tensor through the Gauss-Codazzi equations. Furthermore, we explore the case of a constant electric field, which clearly shows how the bulk B-field pulls on the brane, in a very similar way to how a pulling string induces a point mass \cite{Banerjee:2020wov}. Section \ref{sec: check} is devoted to provide a general proof of the consistency of the model. Finally, results are discussed in section \ref{sec: discussion and outlook}, together with an outline on further steps to accommodate the standard model of particle physics within the dark bubble framework. 
\section{Dark bubble cosmology: un résumé}\label{sec: db_review}

Let us quickly summarise the key features of the dark bubble model that will be relevant to the new results presented in this paper.

\subsection{Hierarchies from an explicit stringy embedding}

According to the dark bubble proposal, our 4D space time can be identified with the surface of an expanding bubble of true vacuum in an unstable five dimensional anti de Sitter ($AdS_5$) space \cite{Banerjee:2018qey}, where, as explained in \cite{Danielsson:2021tyb}, the nucleation of the bubble can be identified with the quantum cosmology of Vilenkin. An explicit stringy realization was proposed in \cite{Danielsson:2022lsl}, where the hierarchy of energy scales was fixed by the top-down construction.\footnote{Other embeddings in non-supersymmetric string theories have been studied in \cite{Basile:2020mpt,Basile:2021vxh}. The type 0'B string \cite{Sagnotti:1995ga, Sagnotti:1996qj} is particularly promising phenomenologically, since it contains charged D3-branes \cite{Dudas:2000sn, Dudas:2001wd}, hence 4d worldvolumes.} We start with the 10D Planck length, which is given by the string coupling and string length in the usual way:
\begin{equation}\label{eq: strtolp}
    l_{10}^4 \sim g_s l_s^4.
\end{equation}
In the following we will, for simplicity, not care about numerical factors. These can be inserted back to get precise relations. Next, we reduce to 5D. The background we consider is a deformation of $AdS_5 \times S_5$. There is no scale separation and the length $L$ sets the size of the $S_5$ as well as of the curvature scale of the non-compact $AdS_5$. The 5D Planck scale is obtained by a simple dimensional reduction on the $S_5$ and is given by
\begin{equation}\label{eq: 10to5}
    l_5^3 \sim l_{10}^8 /L^5.
\end{equation}
The next step is to go down to four dimensions. In a standard dimensional reduction on a compact dimension of size $L$, you would get $l_4^2 \sim l^5/L$. The same result is obtained in the Randall-Sundrum (RS) model on a non-compact AdS-direction with curvature scale $1/L$ \cite{Randall:1999vf}. This is not how it works for the dark bubble. The RS-result for the strength of 4D gravity is due to the bubble being embedded into $AdS_5$ such that it has two insides. The extrinsic curvature, appearing in the junction conditions, have equal signs on both sides. In case of the dark bubble, which nucleates in a pre-existing space time, there is an inside and an outside. As explained in \cite{Banerjee:2020wov}, this leads to a 4D Planck scale given by 
\begin{equation} \label{eq: G4 from G5}
    l_4^2 \sim \frac{ k_- k_+ }{k_- - k_ +} l_5^3,
\end{equation}
where $k_\pm = 1/L_\pm$ with the minus sign referring to the inside, and the plus to the outside. In a dimensional reduction, as well as in the case of RS, the lower dimensional Planck length is {\it smaller} than the higher dimensional one. This means that gravity is {\it weaker} in the lower dimensional theory. Requiring that gravity not to be too strong in the higher dimensions, sets a limit on the size of the extra dimensions; higher the number of large extra dimensions, stricter the limit.

This is no longer true for the dark bubble, which explains many of its unique features. If $k_-$ is close to $k_+$, then $l_4$ can be {\it larger} than $l_5$. In fact, the hierarchy that we will find is $l_{10} \gg l_4 \gg l_5$. Let us look at this in more detail.

As reviewed in \cite{Henriksson:2019zph}, the background $AdS_5 \times S_5$ is dual to a background of $N$ D3-branes. One of these branes can nucleate and start to expand. By studying its junction condition, this was found to correspond to $\left(k_-- k_+\right) \sim k/N \sim \tfrac{1}{NL}$, where we have put $k_- \sim k_+ \sim k$. From this we see that:
\begin{equation}\label{eq: conect planck}
    l_4^2 \sim \frac{N}{L} l_5^3.
\end{equation}
It is a large $N$ what ensures that $l_4 \gg l_5$. As is well known, the scale $L$ of the background is given as $L^2 \sim l_s^2 \sqrt{g_s N}$, which, using (\ref{eq: strtolp}) and (\ref{eq: 10to5}), is equivalent to
\begin{equation}
    l_5^3 \sim \frac{L^3}{N^2}.
\end{equation}
There is only one remaining relation to be fixed, which will determine $N$ in terms of the cosmological constant. The nucleated brane is supposed to be critical and to have a tension equal to the standard value of a D3-brane:
\begin{equation}
    T_3 = \frac{1}{(2 \pi )^3 g_s l_s^4}.
\end{equation}
It is worth pointing out that if all numerical coefficients are restored, it {\it exactly} solves the condition for a critical bubble:
\begin{equation}
    \sigma _c = \frac{3}{8 \pi G_5} (k_- - k_+).
\end{equation}
As argued in \cite{Danielsson:2022lsl}, we expect there to be corrections of order $1/N$ that will {\it reduce} the value of the tension in line with the Weak Gravity Conjecture (WGC)\footnote{This will be discussed and calculated in an upcoming paper \cite{tocome}.} along the lines of the results found in \cite{Bonnefoy:2018tcp, Basile:2021mkd}. This will shift the tension from its critical value, and generate an effective 4D cosmological constant of magnitude
\begin{equation}\label{eq:4d_density}
\rho_\Lambda \sim \frac{1}{g_s l_s^4} \frac{1}{N} \sim \frac{1}{L^4}.
\end{equation}
From this we find that $L \sim 10^{-5} m$, which has the size of the dark dimension discussed in \cite{Montero:2022prj}. This is an intriguing result, but one should keep in mind that in our scenario the scalings do not arise from a hierarchy between a single mesoscopic dimension and the other (scale-separated) ones. Instead, \eqref{eq:4d_density} arises from the peculiar relation between the 5D and 4D Planck scales given by (\ref{eq: G4 from G5}).

Using the previous relations, we can now express the 5D and 10D Planck scales, 4D Hubble length $R_H$, and length scale $L$ of $AdS_5$ in terms of $N$ and the 4D Planck length as:\footnote{If we instead consider corrections to be of order $1/g_s N$, the first relation is changed to $R_H \sim g_s^{1/2} N l_4$.} 
\begin{equation}
    R_H \sim N l_4,\quad  L \sim N^{1/2} l_4, \quad l_{10} \sim N^{1/4} l_4, \quad l_5 \sim N^{-1/6} l_4 . 
\end{equation}
Furthermore, we have $l_s \sim l_{10}/g_s^{1/4}$. This finally fixes $N \sim 10^{60}$. From there we find the non-trivial prediction that the 10D Planck scale should sit at around $10\:  \text{TeV}$, with the string scale just below and a string coupling $g_{s}$ that is less than one but not too small. 

The scalings we have found may be rather general for an embedding meeting the requirements of the dark bubble. The input needed, include the standard relations between scales for a compactification (without scale separation) like $AdS_5 \times S_5$ or some similar warped throat. This, together with the junction conditions for a brane with an inside and an outside, and the tension shifted away from the critical one through a $1/N$-effect, is what leads to the hierarchy.

\subsection{Gravitational waves in dark bubble cosmology}\label{sec: gw_review}

Let us now quickly summarise how gravitational perturbations in the bulk induce gravitational waves (GW) on the bubble's boundary, as studied in \cite{Danielsson:2022fhd}. The four dimensional energy momentum tensor on the brane is induced due to the difference between bulk metrics across the brane, which is captured by Israel's second junction condition,\footnote{Israel's first junction condition requires that the induced metric is continuous across the brane: $h_{ab}^+ = h_{ab}^- \coloneqq h_{ab}$.} given by:
\begin{equation}
    \kappa_{5}S_{ab} = \left.\left[K_{ab}-Kh_{ab}\right]\right|^-_+, \label{eq: junction condition 2}
\end{equation}
where $[A]_{+}^{-}$ stands for the difference between the object $A$ evaluated inside ($-$) and outside $(+)$ the bubble wall. $K_{ab} = \nabla_{\beta} n_{\alpha} \: e^{\alpha}_{a}\: e^{\beta}_{b}$ (with trace $K$) represents the extrinsic curvature and carries information about the bubble's embedding in the bulk geometry.\footnote{The unit normal vector, denoted by $n_{\alpha}$, points in the direction in which the bubble's volume increases.  The tangent vector to the hyper-surface is $e_{a}^{\alpha}$. $h_{ab}$ is the induced metric on the wall. Throughout, bulk coordinates are represented by Greek letter indices, while coordinates on the brane are represented by Latin letter indices.}

We would like to rewrite previous expression in terms of the energy-momentum in the bulk. This can be achieved by inserting the Gauss-Codazzi equation,
\begin{equation}
    R^{(5)}_{\alpha\beta\gamma\delta}e^\alpha_a e^\beta_b e^\gamma_c e^\delta_d = R^{(4)}_{abcd} + K_{ad}K_{bc}-K_{ac}K_{bd}, \label{GaussCodazzi}
\end{equation}
into the junction condition (\ref{eq: junction condition 2}), to find:
\begin{equation}\label{eq: ProjectedEinsteinEqs}
\begin{aligned}
    G_{ab}^{(4)} &= \left(\kappa_4\sigma-3k_+k_-\right)h_{ab}  - \kappa_4 \left(T_{\textrm{brane}}\right)_{ab}+ \frac{k_+k_-}{k_--k_+}\left[\left(\frac{\mathcal{J}_{ab}^+}{k_+}-\frac{\mathcal{J}_{ab}^-}{k_-}\right)-\frac{1}{2}\left(\frac{\mathcal{J}^+}{k_+}-\frac{\mathcal{J}^-}{k_-}\right)h_{ab} \right]\\
    & \quad + \mathcal{O}\left((\Lambda_4/k^2)^2\right),
\end{aligned}
\end{equation}
where $\mathcal{J}_{ab}$ is a tensor defined by:
\begin{equation}
    \mathcal{J}_{ab} = R^{(5)}_{\alpha\beta\gamma\delta}e^\alpha_a e^\beta_b e^\gamma_c e^\delta_d h^{cd}.\label{eq: 5d_tensor}
\end{equation}
and $\mathcal{J}= h^{ab}\mathcal{J}_{ab}$, its trace. $T_{\textrm brane}$ above corresponds to matter on the brane,  and contributes with a negative sign to the energy-momentum tensor in 4D.\footnote{Note that our conventions with metric signature $(-++++)$ and $h_{00}<0$, implies $T_{00}>0$ and $T_0^0<0$ for positive energy.}
We can see from expression (\ref{eq: ProjectedEinsteinEqs}) that it is $\mathcal{J}_{ab}$, the energy-momentum of the bulk, that is in charge of sourcing the four-dimensional geometry. It is important to notice, that even in pure $\text{AdS}_5$, $\mathcal{J}_{ab}$ contributes with $-3k^2 h_{ab}$, which provides a net cosmological constant given by
\begin{equation}
\Lambda_4 =6 k_+ k_- - \kappa_4 \sigma = \kappa_4 \left(\frac{3}{\kappa_5}(k_- -k_+ ) - \sigma \right). 
\end{equation}
Here we see how the tension is matched against the difference in energy between the outside and the inside to yield an effective cosmological constant, always positive. One can then go further and study how different bulk geometries source different energy densities (i.e. dust or radiation) in the four dimensional geometry \cite{Banerjee:2019fzz}. 

Gravitational waves are vacuum solutions of Einstein's equations of motion. They can be described as transverse and traceless (TT) perturbations to a FLRW metric, as:
\begin{equation}
    ds_{\rm{FLRW}}^{2} = a^2(\eta) \left(-d\eta^{2} + \left(\gamma_{ij} +\zeta\: h_{ij}(\eta,x)\right)dx^{i} dx^{j}\right),
\end{equation}
where $\eta$ is the conformal time, $\gamma_{ij}$ is the metric for spatial slices, $h_{ij}$ corresponds to the traceless and transverse perturbations, and $\zeta$ is the perturbation parameter. Assuming a single mode of the perturbation, as $h_{ij}(\eta,x) = h_{4D}(\eta) Y_{ij}(x)$, where $Y_{ij}(x)$ represents spherical harmonics over the closed spatial sections, one finds the time dependent solution to be:
\begin{equation}\label{eq: 4d_wave}
    h_{\rm 4D}\left(\cos(\eta)\right) = \frac{1}{n+1}\cos\left((n+1)\eta\right) +  \sin\eta\sin(n\eta),
\end{equation}
with $n$ labelling the wave number.

The notion of energy of a gravitational wave is a bit tricky. Since it solves the vacuum equation of motion, one could wrongly conclude that they do not carry any energy. The right procedure is to solve the Einstein equations to first order, and then to identify the energy of the first order waves with the second order piece representing the failure of the first order waves to solve the vacuum equations. In \cite{Danielsson:2022fhd} we used this procedure in 4D as well as in 5D. The isotropic, wavelength averaged energy momentum in 4D becomes:\footnote{With coordinates $x^a \equiv \left\{ t,θ,ϕ,ψ \right\}$.}
\begin{equation}
    \langle \tensor{T}{^a_b}\rangle_{\rm iso} = \frac{7}{8\kappa_4}\frac{1}{a^2}\begin{pmatrix}1&0&0&0\\0& 1/3 &0&0\\0&0& 1/3 &0\\0&0&0& 1/3 \end{pmatrix}+\frac{q^2}{4\kappa_4H^2}\frac{1}{a^4}\begin{pmatrix}-1&0&0&0\\0& 1/3 &0&0\\0&0& 1/3 &0\\0&0&0& 1/3 \end{pmatrix}. \label{eq: EM tensor 4D}
\end{equation}
Observe how the first term dilutes as $\sim 1/a^{2}$, behaving like curvature in the Friedmann equations, while the second one, with equation of state $p = -\rho/3$ and dilution $\sim 1/a^{4}$, has a radiation-like behaviour. When the wavelength of the waves is larger than the horizon, it freezes and the only remaining contribution is curvature.

Let us now focus on single modes of TT gravitational perturbations in the five dimensional bulk that will induce the aforementioned four dimensional ones. The partial differential equation to be satisfied, from Einstein's equations is: 
\begin{equation}
        \frac{\partial^2 h}{\partial t^2}-f^2\frac{\partial^2h}{\partial r^2}-\frac{f}{r}\left(2+4k^2r^2+f\right)\frac{\partial h}{\partial r} + \frac{n^2-1}{r^2}fh=0,
        \label{eq: 5D GW equation}
\end{equation}
with $f \equiv f(r)$, the AdS function in global coordinates, and $h \equiv h_{\textrm{5D}}(t,r)$, the wave to solve for. This wave is required to behave like the induced four dimensional one (\ref{eq: 4d_wave}), up to supressed $H/k$ corrections, and to be sourceless ($h_{\textrm{5D}}(t,r) = 0$ when $r\rightarrow 0$). Imposing these boundary conditions, the resulting bulk wave is:
\begin{equation}
    h_{\rm 5D}(t,r) = \frac{(k r)^{n-1}}{\left(1+k^2 r^2\right)^{\frac{n-1}{2}}}\left[\frac{\frac{1}{2}(1+n)(2-n)+k^2 r^2}{(n+1)\left(1+k^2 r^2\right)}\cos\left((n+1)kt\right)+\sin(kt)\sin(nkt)\right]\,.
\end{equation}
with a five dimensional energy momentum:\footnote{With coordinates $x^α \equiv \left\{x^a,r\right\}$.}
 \begin{align}\label{eq: 5d_em}
    &\begin{aligned}\langle\tensor{T}{^\mu_\nu}\rangle_{\rm rad} =  \frac{k^2n^2t^2}{4\kappa_5r^2}\begin{pmatrix}-1&0&0&0&0\\0&1/3&0&0&0\\0&0&1/3&0&0\\0&0&0&1/3&0\\0&0&0&0&0 \end{pmatrix}, 
    &&
    \langle T^\mu{}_\nu \rangle_{\rm flux} =  \frac{n^2}{8\kappa_5r^2}\begin{pmatrix}0&0&0&0&-2tk^{-2}r^{-3}\\0&0&0&0&0\\0&0&0&0&0\\0&0&0&0&0\\2k^2tr&0&0&0&0 \end{pmatrix},\end{aligned}\\
    \nonumber\\
    \nonumber&\langle\tensor{T}{^\mu_\nu}\rangle_{\rm curv} =   \frac{1}{8\kappa_5r^2}\left(7-\frac{n^4}{2k^4r^4}\right)
    \begin{psmallmatrix}1&0&0&0&0\\0&1/3+\mathcal{O}\left(\frac{n^2}{k^2r^2}\right)&0&0&0\\0&0&1/3+\mathcal{O}\left(\frac{n^2}{k^2r^2}\right)&0&0\\0&0&0&1/3+\mathcal{O}\left(\frac{n^2}{k^2r^2}\right)&0\\0&0&0&0&1+\mathcal{O}\left(\frac{n^2}{k^2r^2}\right) \end{psmallmatrix}.
\end{align}
Here, the energy momentum tensor has been averaged and made isotropic, in the same spirit as in the four dimensional case (\ref{eq: EM tensor 4D}). The first line corresponds to gravitational radiation. Note in particular that the non-zero $\{t,r\}$ components of the flux indicate a flow of energy in the direction in which the bubble expands, as expected. The second line corresponds to waves that are frozen as they become larger than the horizon scale.

The very nontrivial check that the dark bubble construction makes sense, is to use the backreacted 5D bulk metric caused by the above energy momentum (\ref{eq: 5d_em}) and calculate its contribution to the extrinsic curvature on the bubble. This metric is given by:
\begin{equation}\label{gwbackreacted5dmetric}
    \d s^2_{\rm back} \approx -\left[1+k^2r^2+ \epsilon^{2}\:\left(q_1-q_2k^2t^2\right)\right] \d t^2 +\left[1+k^2r^2 + \epsilon^{2}\:\left(q_1-q_2k^2t^2\right)\right]^{-1} \d r^2+r^2 \d \Omega_3^2,
\end{equation}
with $|\epsilon|\ll 1$ and $\{q_{i}\}$ coefficients:
\begin{equation}
    q_{1} = -\frac{7}{24},\quad \quad q_{2} = -\frac{q^{2}}{6},\quad \quad q_{3} = \frac{q_{2}}{2} = -\frac{q^{2}}{12}. \label{fixingq}
\end{equation}
This, in turn, gives rise to an effective energy momentum tensor in 4D which, remarkably, is exactly the energy momentum corresponding to the 4D gravitational waves (\ref{eq: EM tensor 4D}).

Let us now see how a similar approach can give rise to electromagnetic waves. Once again, it is crucial to show that the backreaction of the electromagnetic field on the brane induces the correct energy-momentum tensor in the 4D cosmology, effectively changing the sign of the bare contribution $T_\text{brane}$. In the following, we discuss in detail how this is achieved.
\section{Electromagnetic waves in curved spacetime}\label{sec: 4d_em_cosmo}

In this section, we will review concepts of electromagnetism in four dimensional curved space, as a required foundation to understand its higher dimensional embedding. 

\subsection{Maxwell's equations in disguise}

Let us begin by recalling the nature of electromagnetism in a minimally coupled 4D Einstein-Maxwell theory:
\begin{equation}\label{eq:Fμν_action}
    S = \int \d^4 x \sqrt{-g} \left( R - \frac{1}{4g²} F_{μν} F^{μν} \right)\,,
\end{equation}
where $F^{μν}$ is the field strength associated to the $U(1)$ gauge field $A^μ$, and $g$ is the gauge coupling constant. Electromagnetic wave solutions in this background are solutions to sourceless Maxwell's equations, which can be formulated in three equivalent ways:
\begin{enumerate}
    \item The first is the standard Maxwell equations that follow from varying the action
    \begin{equation}\label{eq:maxwell_dF}
        *\d * F = 0 = ∇_β F^{αβ},\quad
        \d F = 0 = ∇_{[μ} F_{νρ]}\,.
    \end{equation}
    \item Maxwell's equations above can be combined by taking a covariant derivative of the second equation and using the first. This gives a wave equation for the electromagnetic tensor
    \begin{equation}\label{eq:wave_fab}
    ∇^α ∇_α F_{ab} \equiv ∇²F_{ab} = -2R_{abcd}F^{cd} + R_a{}^c F_{cb} + F_a{}^c R_{cb}\,.
    \end{equation}
    This highlights an interesting feature of curved spacetime: electromagnetic waves couple to the spacetime curvature. Consequently, curvature can act as a source for electromagnetic waves. Recall that $F_{ab}$ can be decomposed along the world line of an observer (see \cite{Tsagas:2004kv, Mavrogiannis:2021qtc} for further details)
    \begin{equation}
    F_{ab} = 2u_{[a]}E_{b]} + \epsilon_{abc} B^c\,,
    \end{equation}
    where $u^a \coloneqq \d x^a/\d τ$ is the 4-velocity tangent to the observer's worldline.
This implies that the electric and magnetic fields experienced by the observer are given by the projections $E_a =F_{ab}u^b$, and $H_a =\epsilon_{abc} F^{bc}/2$.
Projecting \eqref{eq:wave_fab} along $u^a$ gives the wave equation for the electric field $E^a$, and its dual gives the wave equation for the magnetic field.

\item Yet another way to reformulate Maxwell's equations is to write them for the gauge field instead of the field strength
\begin{equation}\label{eq: dRham_eq}
     ∇_β ∇^β A^α - ∇^α ∇_β A^β - R^α{}_β A^β = 0\,.
\end{equation}
The curvature term comes from commuting the covariant derivatives. One can use the gauge freedom in the system to choose Lorenz gauge: $∇_α A^α = 0$. With this choice, the left hand side becomes the de Rham operator $Δ_{\textrm{(dR)}}$, which is the ordinary Laplacian supplemented with the Ricci curvature. The resulting wave equation is also referred to as the de Rham equation (see for example \cite{deRham})
\begin{equation}
    Δ_{\textrm{(dR)}} A^α \coloneqq ∇_β ∇^β A^α - R^α{}_β A^β = 0\,.
\end{equation}
\end{enumerate}
The three formulations above are equivalent in content, and one can choose to solve either of them. The second and third formulations are inherently second order in the field strength and the gauge field respectively, while the first formulation in \eqref{eq:maxwell_dF} involves only a single derivative on $F^{μν}$. Additionally, the third formulation is simple only in the Lorenz gauge, which will need to be generalized for higher form fields that we will be interested in solving wave equations for in the next section. The best bet is therefore to use the first formulation, which is what we will do.

\subsection{Electromagnetic waves in a conformally flat de Sitter universe}
Let us solve for EM waves in a homogeneous isotropic conformally flat universe whose metric is given by
\begin{equation}
    \d s² = a(\eta)² (-\d \eta² + \d x_i²)\,,
\end{equation}
with a scale factor $a(\eta)=\left(H \eta\right)^{-1}$, this is a patch of conformally flat de Sitter spacetime.
Decomposing $F^{μν}$ in a basis of plane waves, let us make the ansatz $F^{μν} = e^{i n_λ x^λ} f^{μν}(\eta)$, where $n_{\lambda}$ represents the wave vector. To further simplify the problem, we will consider waves propagating along one of the three isotropic space directions (say $z$). The full isotropic wave can be obtained by averaging the solution over all space directions.
Equations \eqref{eq:maxwell_dF} imply (for $i=x, y$)
\begin{equation}
    f^{t,z} =
    \eta \dot{f}^{\eta,i} -4 f^{\eta,i} - i n\,\eta f^{i,z} = 
    -\eta \dot{f}^{i,z} + 4 f^{i,z} + i n \,\eta f^{\eta,i} = 0\,,
\end{equation}
where a dot denotes a conformal time derivative. This has a very simple solution
\begin{equation}\label{eq:sol_wave_frw}
\begin{aligned}
    F^{\eta,x} &= \frac{e^{inz}}{a^4} \left( c_1 \cos(n \, \eta) + c_2 \sin(n \, \eta)  \right),\quad
    F^{x,z} = \frac{e^{inz}}{a^4}i \left( c_2 \cos(n \, \eta) - c_1 \sin(n \, \eta)  \right)\,\\
    F^{\eta,y} &=  \frac{e^{inz}}{a^4} \left( c_3 \cos(n \, \eta) + c_4 \sin(n \, \eta)  \right),\quad
    F^{y,z} = \frac{e^{inz}}{a^4} i \left( c_4 \cos(n \, \eta) - c_3 \sin(n \, \eta)  \right).
\end{aligned}
\end{equation}

As a demonstration of another way to solve the problem, let us use the third formulation above (\ref{eq: dRham_eq}) and solve the wave equation for the gauge field $A^μ$. Making an ansatz for the gauge field $A^μ = \left(0, A^i(\eta) e^{inz}\right)$, the Lorenz gauge condition gives $∇_μ A^μ = ine^{ikz} A^3 = 0 ⇒ A^3 = 0$. With this, de Rham's equation gives
\begin{equation}
    \eta^{2} \left(\ddot{A}^i +n² A^i \right) - 4 \eta \dot{A}^i + 6 A^i = 0,\quad \textrm{for }i\in{x,y} \textrm{ and 0 otherwise}.
\end{equation}
With the redefinition $A^i(\eta) \coloneqq B^i(\eta) \, \eta^{2}$, this becomes a simple wave equation $\ddot{B}^i +n² B^i = 0$, which gives the same solutions as \eqref{eq:sol_wave_frw} above. 

The energy momentum tensor carried by the electromagnetic wave is given by
\begin{equation}
    T^a{}_b = F^{ac}F_{bc}-\frac{1}{4}δ^a{}_b F^{cd}F_{cd}\,.
\end{equation}
A uniform background of electromagnetic radiation requires to average overall possible phases and directions of the waves with different polarizations. This results in an isotropic energy momentum tensor of the form:
\begin{equation}\label{eq: avg_4d_tensor}
   \langle T^a{}_b\rangle_{\rm iso} = \frac{\mathcal{E}^{2}}{2 a^4}
    \begin{pmatrix}
    -1&0&0&0\\
    0&\frac{1}{3}&0&0\\
    0&0&\frac{1}{3}&0\\
    0&0&0&\frac{1}{3}\\
    \end{pmatrix},
\end{equation}
where $\mathcal{E}$ accounts for the amplitude of the wave, encoded by the $c_i$ in eq \eqref{eq:sol_wave_frw}. 
This section foreshadows the next, where we will solve for vacuum waves of the Kalb-Ramond 2-form field $B^{μν}$. If we were to use the third formulation to solve for the waves there, we would first need to find an analog of the Lorenz gauge for the problem. A simple guess would be to use $∇_μ B^{μν} = 0$. However, this leaves behind extra curvature terms that do not result in a de Rham wave equation, and makes it very complicated to solve. Another complication arises when one tries to find a suitable ansatz for the $B^{μν}$ field. A natural guess is to make an ansatz that reduces on the shell to the 4D $A^μ$ found above. However, there are multiple ways of doing this, which complicates the problem. One could choose a better gauge, or make a better ansatz, but even then, one has to solve a cumbersome second order PDE, which carries its own set of challenges. For the problem that we are interested in, we will therefore use the first formulation to solve directly for the 3-form field strength $H_{μνρ}$ instead.

\section{Fiat lux in obscura bulla\footnote{In English, Let there be light in the dark bubble.}}\label{sec: uplift}

In this section, we are going to study how gauge fields in the $AdS_{5}$ bulk, which also enjoy a ten dimensional supergravity interpretation, can give rise to their four dimensional counterparts on the boundary of the bubble.

\subsection{The uplift of electromagnetic waves}

We will start by exploring the Dirac-Born-Infeld (DBI) action for a D3-brane, given by\footnote{Note the change in notation in this section where Greek letters are used to label 4D indices on the brane.} 
\begin{equation}\label{eq: DBI action}
S_{D3} = -T_3 \int \d^4 x \sqrt{- \det (g_{4}+ \tau \mathcal{F})},
\end{equation}
with $\tau = 2 \pi \alpha'$, where $\alpha ' \equiv l_s²$ is the string tension. Here and in the following we suppress indices in the determinants. In string theory there is also a coupling to bulk RR-flux, but we will not consider this here. There is an immediate and obvious candidate for the EM-waves: the gauge field $\cal{F}_{\mu \nu}$. Let us now explicitly expand (\ref{eq: DBI action}) up to second order for small $\alpha'$, which results in:
\begin{equation}\label{eq: expand_dbi}
   S_{D3} = - \int \d^4 x \sqrt{- \det g_{4}} \left(T_3 + \underbrace{T_3 \,\tfrac{\tau^{2}}{4}}_{1/\left(4 g^{2}\right)} \mathcal{F}_{\mu\nu} \mathcal{F}^{\mu\nu}+ \mathcal{O}\left(\mathcal{F}\mathcal{F}\right)^{2}\right),
\end{equation}
where $g^{2} = 2 \pi g_{s}$ is the gauge coupling. The second piece in eq (\ref{eq: expand_dbi}) can then be identified with the action of electromagnetism in four dimensions (\ref{eq:Fμν_action}). However, contrary to the case of other kinds of brane world constructions, this is far from the whole story. The first puzzle to solve is that the energy density of such a gauge field would seem to {\it add} to the energy of the tension, and therefore, recalling, e.g. \cite{Banerjee:2019fzz}, it would give a {\it negative} contribution to the 4D energy momentum. Nevertheless, as explained in \cite{Danielsson:2022fhd}, we also need to take into account the back reaction from the bulk. We will come back to this at the end of the section.

\begin{figure}[ht!]
    \centering
    \includegraphics[width=0.5\textwidth]{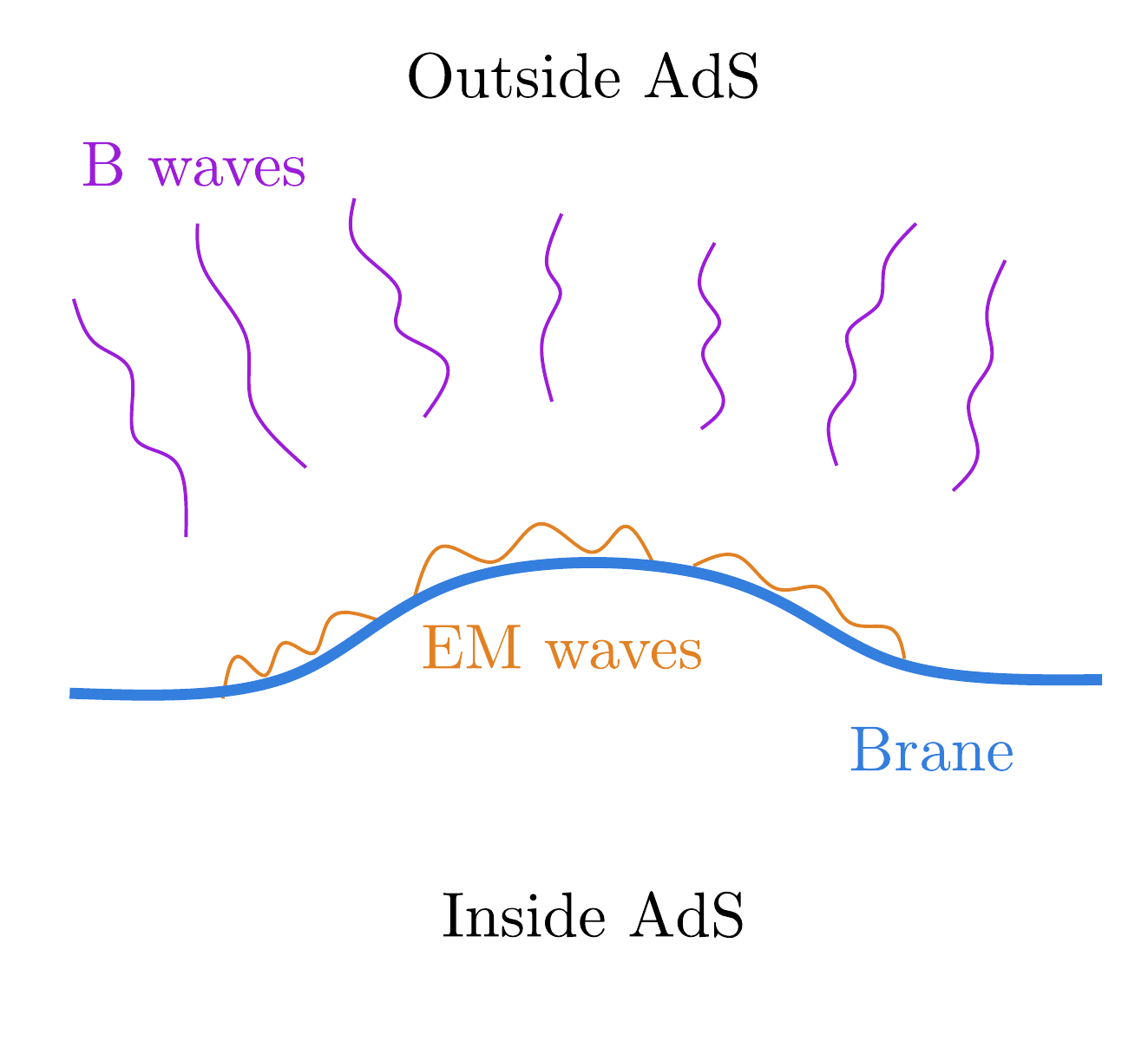}
    \caption{A sketch of the dark bubble carrying electromagnetic waves sourcing the B-field in the bulk. The junction conditions imply a jump in the B-field encoded in eq \eqref{eq:Hjump}.}
    \label{fig:B-field_bubble}
\end{figure}

As it is well known, the gauge field on the brane can be written as:
\begin{equation} \label{eq: ffb}
\tau \:{\cal F}_{\mu \nu}= \tau \: F_{\mu \nu}+ B_{\mu \nu}.
\end{equation}
It is only ${\cal F}$ that is gauge invariant, while $F$ needs to shift when there is a change of gauge for the B-field. The bulk field strength $H=dB$ is of course also gauge invariant. By varying the DBI action (\ref{eq: DBI action}) with respect to $B$, we get a source term for $H$ given simply by ${\cal F}$. Since we are working with a shell with co-dimension one, this means that there is a jump in $H$ across the shell. We assume that $H$ vanishes inside the shell and that the non-vanishing value outside is purely due to the presence of the electromagnetic field on the brane. To be more precise, the bulk equations in $AdS_5$ that we need to solve, are obtained from the following bulk action, with the DBI-contribution included: 
\begin{equation}\label{eq: full_bulk_action}
    S_5= \frac{1}{2\kappa_5} \int \d^5 x \sqrt{ - \det \left(g_5\right)} \left( R - \frac{1}{12g_s}H^2 \right) - T_{3} \int \d^{5}x \:\delta(r-a[\eta])\: \sqrt{-\det\left(g_{4} + \tau \cal{F}\right)}.
\end{equation}
Here we work in the 5D Einstein frame, where the dilaton coupling of the B-field is fixed to $e^{-\phi} = g_s^{-1}$, and $\sqrt{-\det \left(g_5\right)}= \frac{1}{kr} \sqrt{-\det \left(g_4\right)}$, with the bulk metric of the form:
\begin{equation}
    \d s_5^2 = k^2 r^2 \left( -\d t^2 + \d x^2 + \d y^2 + \d z^2\right)+\frac{\d r^2}{k²r^2}.
\end{equation}
Note that the conventions are such that ${\tau \cal F}$  and $B$ are dimensionless. We assume that internal (geometric) moduli and the dilaton are stabilized. If we expand for small $\alpha'$, as we did in expression (\ref{eq: expand_dbi}) and then vary the action with respect to $B$, we find the equations of motion
\begin{equation}
     \partial_r H^{r\mu \nu} = \frac{2 \kappa_{5} \:kr}{\alpha'\: \pi^{2} } \:{\cal F}^{\mu \nu} \:\delta \left(r-a[\eta]\right),
\end{equation}
Integrating across the brane at $r=a(η)$ gives
\begin{equation}\label{eq:Hjump}
    \Delta H^{r\mu \nu} \big|_{r=a} = H^{rμν} \big|_{r=a} = \frac{2 \kappa_{5} \:ka}{\alpha' \:\pi^{2} }\:\: {\cal F}^{\mu \nu} \big|_{r=a}\,.
\end{equation}
Let us now understand how the four and five dimensional field strengths relate to one another. We first connect the field strength $\mathcal{F}$ in the DBI piece of expression (\ref{eq: full_bulk_action}) and that from Maxwell in eq (\ref{eq:Fμν_action}). To be specific, we focus on electromagnetic waves propagating along $z$, with the electric field polarized along $x$, and the magnetic field polarized along $y$. Given a wave propagating in the $z$ direction, and $\mathbf{E,B}$ respectively polarised along $\{x,y\}$ directions, we have
\begin{equation}
    {\cal F}^{t x} = \frac{{\cal E}(t,z)}{k^2 a^4}, \quad  {\cal F}^{x z} =  \frac{{\cal E}(t,z)}{k^2 a^4} , 
\end{equation}
where $\cal E$ is a dimensional function that determines the amplitude of the wave.\footnote{Here we have used {\it conformal} coordinates with ${\cal F}^{\mu\nu} \sim 1/ k^2a^4$ and ${\cal F}_{\mu\nu} \sim k^2$ as in section 3.1. With {\it proper} time and length given by $\d x^\mu_p = ka \d x^\mu$ we would instead have ${\cal F}^{\mu\nu} \sim {\cal F}_{\mu\nu} \sim 1/a^2$.} 
With this information, we can then identify the behaviour of the $H$ field strength on the brane, assuming there is no $H$-field turned on inside of the bubble. Imposing this boundary condition on expression (\ref{eq:Hjump}), gives us:
\begin{equation}\label{eq : H_to_F}
    \begin{aligned}
        H^{rtx}\big|_{r=a} &\equiv h_{1} (t,a,z) = \frac{2 \kappa_{5} \:ka}{\alpha'\: \pi^{2} } {\cal F}^{tx} = \frac{2 \kappa_{5}}{\alpha'\: \pi^{2} k}\frac{{\cal E}(t,z)}{a^3},\\
        H^{rxz}\big|_{r=a} &\equiv h_{2} (t,a,z) = \frac{2 \kappa_{5} \:ka}{\alpha'\: \pi^{2} } {\cal F}^{xz} = \frac{2 \kappa_{5}}{\alpha'\: \pi^{2} k}\frac{{\cal E}(t,z)}{a^3},
    \end{aligned}
\end{equation}
In figure \ref{fig:B-field_bubble} we schematically see how the electromagnetic waves on the brane source the H-field in the bulk above the brane. 

The $H$ field must solve the equation of motion, as well as the Bianchi identities away from the brane.\footnote{Note that we assume that the dilaton as well as any other moduli of the background are fixed.} We get:
\begin{equation}\label{eq: eom_bianchi}
    \begin{aligned}
       * \d * H = 0 &⇒ \nabla_{\alpha}H^{\alpha \beta \gamma} = 0\,,\\
      * \d H=0 &⇒ \epsilon^{\alpha\mu\nu\lambda}_{\quad\quad  \kappa} \partial_{\alpha}H_{\mu\nu\lambda} = 0\,,
    \end{aligned}
\end{equation}
where $*$ is the 5D Hodge star and $H$ is the 3-form corresponding to $H$ above.
As we will see, it is not enough to turn on these components of the bulk field. Any non-trivial time dependence also requires the presence of $H^{txz} \equiv h_{3} (t,r,z)$. 

For the AdS-background, in the limit of flat 4D space time, the equations of motion become:
\begin{equation}\label{eq: EOM_wave}
    \begin{aligned}
        \left( \partial_r+\frac{3}{r} \right) h_1+\partial_z h_3&=0,\\
        \left( \partial_r+\frac{3}{r} \right) h_2+\partial_t h_3&=0,\\
        \partial_z h_2-\partial_t h_1&=0,
    \end{aligned}
\end{equation}
while the Bianchi identity becomes 
\begin{equation}\label{eq: Bianchi}
    r^2(\partial _t h_2 - \partial _z h_1) +\: k^{4}\:\partial_r \left(r^6 h_3\right)=0
\end{equation}
A trivial solution is $h_1=h_2=(kr)^{-3} \sin(kn(t+z))$ with $h_3=0$, and 
the momentum of the wave given by $n/r$. Using eq \eqref{eq: em_tensor_derived}, this trivial solution gives an energy momentum tensor of the form:
\begin{equation}\label{eq: trivial}
    T^{\mu}{}_{\nu} \sim \frac{1}{r^{4}} \:\text{diag}\left(-1, 1, 1, 1, 0\right),
\end{equation}
which does not correspond to the expected field strength in the bulk. To be more precise, with such an energy momentum tensor the back reacted bulk metric will contain a logarithmic dependence on the scale factor. If the brane were to move in this background, the 4D observer would note an unphysical change in the energy density on the brane, in addition to the familiar $1/a^4$. To compensate for this, we must have an explicit time dependence of the bulk metric, corresponding to time dependent flux of energy in the radial direction, which corresponds to a non vanishing $T^{t}_{\:\: r}$ component. 

Let us now think of a more general ansatz for the $H$ field strength, accounting for the aforementioned $H^{txz}$ component that represents such a flux. We make the ansatz:
\begin{equation}\label{eq: ansatz}
        h_1= f_1 (t, z, x)\,k^{-2}r^{-3},\, \qquad
        h_2= f_2 (t, z, x)k^{-2}r^{-3},\, \qquad 
        h_3= f_3 (t, z, x)\,k^{-3}r^{-4},\,
\end{equation}
with $x=2 k^{3}t r²n^{-1}$ and 
\begin{equation}\label{eq:dots}
    f_{i} (t,z,x) \equiv \alpha_{i}(t,x) \sin\left(kn \left(t+z\right) \right) + \beta_{i}(t,x) \cos\left(kn \left(t+z\right) \right),
\end{equation}
where the functions $\alpha_i$ and $\beta_i$ are dimensionless. The reason for an ansatz of this form is the following. We note that  on the brane (i.e. $\eta = kt = -\left(Hr\right)^{-1}$):
\begin{equation}
    \frac{1}{knt} \sim \frac{rH}{n} \sim \frac{H}{p}\,,
\end{equation}
where $p\sim n/r$ is the energy of the quanta. Usually, $p \gg H$ and hence $1/\left(knt\right) \ll 1$, is a natural variable for expansion. We also note that, again on the brane:
\begin{equation}
    x=\frac{n}{2 k^{3}tr²} \sim \frac{nH}{k^{2}r} \sim \frac{pH}{k²} \sim \frac{H}{k} \cdot \frac{p}{k} \sim \frac{p}{M_{pl}}\,,
\end{equation}
where in the last step, we have used relations from \cite{Danielsson:2022lsl}, where $1/H \sim N l_4$ and $1/k \sim N^{1/2} l_4$ and $M_{pl}$ is the Planck mass in 4D. It is amusing to see how these specific results guarantee that the corrections remain small until $p$ reaches the Planck scale. To summarize, we conclude that the corrections are small, provided that $H \ll p \ll M_{pl}$, which is also the relevant regime for us.

With this ansatz, we find the solution to the equations of motion and the Bianchi identity, to order $(knt)^{-1}$ but all orders in $x$, given by:
\begin{eqnarray}
    h_1 (t,z,x)=\frac{\alpha}{k^{2}r^3} \sin\left(kn \left(t+z\right)\right) + \frac{2\beta}{k^{2}r^3}  \left[ \sin\left( \tfrac{x}{2}+ kn \left(t+z\right) \right) - \sin\left( kn \left(t+z\right) \right)\right],\\
    h_2(t,z,x)=h_1(t,z,x) -\frac{\beta x}{k^{3} n t r^3} \cos\tfrac{x}{2}\sin\left(kn \left(t+z\right)\right),\\
    h_3(t,z,x)= \frac{2\beta x}{n k^{3}r^4} \sin\left( \tfrac{x}{2}+ kn \left(t+z\right) \right),
\end{eqnarray}
where
$\alpha$ and $\beta$ are now constants. We note that it is only $h_2$ that needs a correction at $(knt)^{-1}$.

\subsection{Energy-momentum tensor in the bulk}

Varying the bulk part of the action (\ref{eq: full_bulk_action}) with respect to $g_{\mu\nu}$, yields the energy-momentum tensor for the electromagnetic fields as:
\begin{equation}\label{eq: em_tensor_derived}
    T_{\tau \omega} = \frac{1}{\kappa_5}\left(-\frac{1}{2} g_{\tau \omega} H^{2} + 3\:H_{\tau \alpha \beta}H_{\omega}^{\:\:\:\alpha \beta}\right).
\end{equation}
So far, our previous computations have been for a single electromagnetic wave traveling in the $z$ direction, with electric and magnetic fields accordingly polarised along $\{x,y\}$ directions. A uniform background of electromagnetic waves can be realised by accounting for all ingoing and outgoing waves along some given direction. Furthermore, one should average over waves traveling isotropically (along $\{x,y,z\}$) with any possible polarisation. This results in the following isotropic energy-momentum tensor: 
\begin{equation}\label{eq: EM_tensor_H}
    \langle\tensor{T}{^\mu_\nu}\rangle_{\rm iso} = \langle\tensor{T}{^\mu_\nu}\rangle_{\rm rad}+ \langle\tensor{T}{^\mu_\nu}\rangle_{\rm flux},
\end{equation}
where the radiation and flux contributions to the 5D energy-momentum tensor are given by
\begin{equation}
\langle\tensor{T}{^\mu_\nu}\rangle_{\rm rad} = \frac{3 \alpha^{2}}{\kappa_5 \, k^{2} r^{4}}
\begin{pmatrix}
-1&0&0&0&0\\
0&\frac{1}{3}&0&0&0\\
0&0&\frac{1}{3}&0&0\\
0&0&0&\frac{1}{3}&0\\
0&0&0&0&0
\end{pmatrix},
\qquad\quad
\langle\tensor{T}{^\mu_\nu}\rangle_{\rm flux} =
\frac{3 \alpha^{2}}{\kappa_5 \, k^{2} r^{4} \, t}
\begin{pmatrix}
0&0&0&0&\frac{-\beta}{ \alpha \, k^{4} r^3}\\
0&0&0&0&0\\
0&0&0&0&0\\
0&0&0&0&0\\
\frac{\beta \,r}{\alpha}&0&0&0&0
\end{pmatrix}.
\end{equation}
Off-diagonal terms in the flux correspond to the energy flow that we anticipated,\footnote{If we were to lower the upper index, the energy momentum tensor is of course symmetric.} analogous to energy flow along the radial direction in eq. \eqref{eq: 5d_em}. Covariant conservation of this energy momentum tensor is guaranteed, as each component $h_{i}$ of the bulk field strength $H$ solve both their equation of motion (\ref{eq: EOM_wave}) and the Bianchi identity (\ref{eq: Bianchi}).

In the next subsection we will calculate the backreaction and move on to matching parameters with the 4D dark bubble cosmology.

\subsection{Backreaction}

As already hinted in eq (\ref{eq: trivial}), in the case where there is no net flux along $tr$ direction, the metric will have the general form of an AdS-Schwarzschild-like one, where $f(r) = 1 + k^2 r^2-1/r^2$ should be replaced by $\hat{f}(r) = 1 + k^2 r^2 -\log r/r^2$. In 4D, this would correspond to a logarithmic increase of energy with time on the brane, compared to the expected $1/a^4$ decay as the universe expands. What happens is that the expanding bubble scoops up matter from bulk that is added to the brane. To avoid this, we need the bulk matter to expand up along the throat of AdS, just as it does in the case of GW \cite{Danielsson:2022fhd}. This is precisely what we have achieved above.  

The energy momentum tensor in the bulk that does the job has $\left(\log r\right)/r^2 \rightarrow \left(\log tr\right)/r^2$. Since $kt=-1/\left(Hr\right)$ on the bubble this give the correct behaviour in 4D.\footnote{It is amusing to compare with the case of gravitational waves. There, $1/r^2 \rightarrow t^2 r^2/r^2$. Presumably this has to do with the fact that electromagnetism is scale invariant and renormalizable, while this is not the case for gravity. The cosmological expansion mirrors the behaviour of renormalization group flows, given the holographic nature of the direction along the throat of the AdS.} To be more precise, the 5D metric (in the flat 4D limit) is given by
\begin{equation}\label{EMbackreacted5dmetric}
   \d s^2_{\rm back} \approx -f_1(t,r) \d t^{2} + f_2(t,r)^{-1} \d r^{2} +r^{2} \d \Omega_{3}^{2} ,
\end{equation}
where
\begin{equation}
    f_i(t,r) \equiv k^{2}r^{2}-\frac{\epsilon^{2}\left(\log (-\xi k^{2} t r)+q_i\right)}{k^{2}r^{2}}
\end{equation}
with $\{q_{i}\}$ undetermined numbers. This line invariant yields an energy momentum tensor of the form:
\begin{equation}\label{eq: EM_tensor_back}
\langle\tensor{T}{^\mu_\nu}\rangle_{\rm back} = \frac{3 \: \epsilon^2}{2 \, \kappa_5 \, k^{2}\, r^4}
\begin{pmatrix}
-1&0&0&0&\frac{-1}{k^{4} r^3 t}\\
0&\frac{1}{3}&0&0&0\\
0&0&\frac{1}{3}&0&0\\
0&0&0&\frac{1}{3}&0\\
\frac{r}{t}&0&0&0&0 
\end{pmatrix},
\end{equation}
provided that $q_1-q_2 =1/4$. Note that $\xi$ is a free parameter that does not affect the energy momentum tensor. Tuning it corresponds to changing  a piece of the metric that is the vacuum AdS-Schwarzschild background. Fixing it will play an important role later on.

Let us now compare both energy momentum tensor (\ref{eq: EM_tensor_H}) and (\ref{eq: EM_tensor_back}) to relate the parameters $\alpha, \beta, \epsilon$. It is easy to see that:
\begin{equation}\label{eq: identification_epsilon}
    \alpha = \beta =\frac{\epsilon}{\sqrt{2}} .
\end{equation}
We can also express these coefficients in terms of the amplitude of the 4D electromagnetic wave. Comparing to eq (\ref{eq : H_to_F}) we can read off that:
\begin{equation}\label{eq : alpha_to_e}
    \alpha= \frac{2\kappa_5 k}{\alpha ' \pi^2} \mathcal{E}.
\end{equation}
Finally, the covariant conservation of the energy momentum tensor (\ref{eq: EM_tensor_back}) is still respected, as it coincides with that derived for the $H$ field in eq (\ref{eq: EM_tensor_H}) and it obeys the Einstein equation.

\subsection{Checking the induced 4D energy momentum tensor on the bubble}

Let us now study how the backreaction of the bulk geometry, due to the presence of $H$ field strength, changes the motion of the bubble wall, and hence the induced energy momentum tensor. One can equivalently arrive to the same result by computing Israel's second junction condition (\ref{eq: junction condition 2}) or the Gauss-Codazzi equation (\ref{eq: ProjectedEinsteinEqs}). There will two main contributions to the 4D energy momentum tensor: 

\begin{enumerate}
    \item A {\it positive} contribution coming from the extrinsic curvature.
    \item A piece that comes from the gauge-field \textbf{on} the brane, which \textbf{adds} to the tension and thus gives a {\it negative} contribution.
\end{enumerate}

Israel's second junction condition (\ref{eq: junction condition 2}) yields an induced Friedmann equation of the form:
\begin{equation}\label{eq: induced_friedmann}
    \begin{split}
        \left(\frac{\dot{a}}{a}\right)^{2} \delta^{a}_{b} = \frac{\Lambda_{4}}{3}\delta^{a}_{b}+ \:&\epsilon^{2} \, \frac{\kappa_{4}}{6 \kappa_{5} k^{3} a^{4}} \left(3 \delta^{a}_{0}\delta^{0}_{b}
        - \delta^{a}_{i} \delta^{i}_{b}\right) \left(q_{2} + \log\left[- \xi \frac{k_{+}}{H}\right]\right)\\
        &+\frac{\kappa_{4}}{3}\left( \mathcal{F}^{ac}\mathcal{F}_{bc}-\frac{1}{4}\,\delta^{a}_{b}\mathcal{F}_{ij}\mathcal{F}^{ij} \right).
    \end{split}
\end{equation}
Note that the left hand side (LHS) is the usual geometrical evolution of the scale factor for late time cosmologies (i.e. there is no contribution of the curvature term) in Friedmann equations, while the right hand side (RHS) corresponds to the energy momentum tensor associated with the induced cosmology. For indices $a=b=0$ we obtain the first Friedmann equation, and one notes that the gauge field on the brane, in the second line, gives a {\it negative} contribution to the energy density, just like the brane tension does. Recall that our conventions are such that $T^0{}_0 = -\rho$. This needs to be corrected to a positive contribution by the first term due to the extrinsic curvature.

We still have not determined the free parameters $\xi$ and $q_2$, which are degenerate and can be shifted into each other. It corresponds to the fact that we can always add a piece of dark radiation by adding a 5D black hole in the background. This was already explored in \cite{Danielsson:2022fhd}. Let us now impose the absence of such contribution, which restricts the RHS of eq (\ref{eq: induced_friedmann}) above to be that of four dimensional electromagnetism in an expanding cosmology, as given in eq (\ref{eq: avg_4d_tensor}). Furthermore, for simplicity, we choose the free parameter $\xi$ so that $-\xi \frac{k}{H}=1$. This implies:
\begin{equation}
    \epsilon^{2}\left( \frac{q_{2}}{2 a^{4} k_{+}^{3} \kappa_{5}}\right) \left(3 \delta^{a}_{0}\delta^{0}_{b} - \delta^{a}_{i} \delta^{i}_{b}\right) + \underbrace{\left(\mathcal{F}^{ac}\mathcal{F}_{bc}-\tfrac{1}{4}\,\delta^{a}_{b}\mathcal{F}_{ij}\mathcal{F}^{ij}\right)}_{\langle T^{a}_{\:\: b}\rangle_{\rm iso}}\equiv - \langle T^{a}_{\:\: b}\rangle_{\rm iso}.
\end{equation}
Making use of relations (\ref{eq: identification_epsilon}), (\ref{eq : alpha_to_e}) and (\ref{eq: conect planck}) one can fix $q_{2}$ to be:
\begin{equation}
    q_{2} = \frac{k_{+}^{2} M_{pl}^{2}}{2^{7}\, T_{D3} \, g_{s}}\, N \, \pi,
\end{equation}
where $N$ is the number of $D3$ branes in the 10D background, as explained in section \ref{sec: db_review}.\footnote{Using the scaling relations, one finds that $q_2 \sim N/g_s$.}

This is all analogous to the case where we {\it only} have dark radiation associated with a black hole with mass $M$ in the bulk, instead of the electromagnetic waves. As already discussed in \cite{Banerjee:2022myh}, the 4D energy momentum tensor has two contributions yielding
\begin{equation}
\rho =\frac{M_+}{k_+ a^4} -\frac{M_-}{k_- a^4} ,
\end{equation}
with $M_- = M_+ = M$. If we now \textbf{move} the mass from the center of the bubble and deposit it evenly onto the brane, without changing the exterior metric this becomes
\begin{equation}
\rho =\frac{M_+}{k_+ a^4} -\rho _\mathrm{brane}\,.
\end{equation}
So, the matter on the brane actually does give a \textit{negative} contribution, but it is overcompensated by the contribution from the extrinsic curvature of the exterior. This is nothing but the gravitational back reaction from the bulk. This is exactly what happens for the electromagnetic field. As we have seen, the back reaction happens through the $H$-field that sources the term $q_2$ in the metric that plays the role of $M_+$. We will come back to a more general discussion of this structure in the next section.

It is important to note that electromagnetic radiation is realized in a very different way from the dark radiation. It is accompanied by a non-trivial $H$-field, with a non-vanishing energy density in the bulk. This yields a different bulk geometry. In order to further highlight the role of the B-field in determining the correct energy-momentum tensor on the brane, we will now consider electrostatic field configurations.

\subsection{Un pliage \'electrique \footnote{En anglais, an electric bending.}}

Let us consider a constant electric field $\cal E$, pointing in the $z$-direction. For simplicity, we ignore the expansion of the universe. Also, such a constant electric field cannot be supported over an extended region. In practice, it could be approximately realized as a piece of an electric field far away from a localized source. In this case, the solution of eqs \eqref{eq: EOM_wave} are very simple, and we find 
\begin{equation}
    H^{rtx}\big|_{r=a} \equiv h_1=\frac{\gamma}{k^2 r^3}, \quad H^{rxz}\big|_{r=a} \equiv h_2=0, \quad H^{txz}\big|_{r=a} \equiv h_3=0 ,
\end{equation}
where $\gamma \sim {\cal E}$ is a dimensionless constant.
This sources a 5D energy-momentum tensor
\begin{equation}
\langle T ^\mu{}_\nu\rangle \approx \frac{\gamma^2}{k^2 \kappa_5}\begin{pmatrix}-1/r^4&0&0&0&0\\0&1/r^4&0&0&0\\0&0&1/r^4&0&0\\0&0&0&1/r^4&0\\ 0&0&0&0&-1/r^4 \end{pmatrix}\,,
\end{equation}
which leads to the following back reaction in the bulk:
    \begin{equation}\label{eq: constE backreact}
        \d s^{2}_{\rm back} \approx A(r) \left(- \d t^{2}+ \d z^{2}\right) + \frac{1}{A(r)^{2}} \d r^{2}  + B(r) (\d x^{2} + \d y^{2}),
    \end{equation}
with 
\begin{align}
    A(r) & = (k r)^{2} -\frac{2}{3} \gamma^{2} \frac{\log(k \zeta r)}{(kr)^{2}},\\
    B(r) & = (k r)^{2}+ \gamma^{2} \frac{\log(k \zeta r)}{(kr)^{2}}\,,
\end{align}
where $\zeta$ is a constant.
The next step is to find the embedding of the brane that gives the correct induced metric in 4D corresponding to a constant electric field. To do this, we must construct the constant electric field as a local approximation within a setup containing an explicit, physical source. Let us use the 4D metric of a point source, where we are far enough from the source that the electric field can be considered constant in a big enough region. We will not take into account any gravitational forces, so $1 \gg Q^2/r^2 \gg M/r$. (This is the relevant situation for any currently realistic experiment.) The induced 4D metric we need on our embedded brane is therefore of the form
\begin{align} \label{eq: RN}
    \d s^2&= -\left(1+\frac{Q^2}{\rho^2} \right) \d t^2 +\left(1+\frac{Q^2}{\rho^2}\right)^{-1} \d ρ^2 + \rho^2 \d\Omega _2 \nonumber \\
    &\rightarrow -\left(1+\frac{Q^2}{\rho^2} \right) \d t^2 +\left(1+\frac{Q^2}{\rho^2} \right) \d\tilde{\rho}^2 + \rho^2 \d\Omega _2 
    \nonumber \\
    &\rightarrow -\left(1+\frac{Q^2}{z^2} \right) \d t^2 +\left(1+\frac{Q^2}{z^2} \right) \d\tilde{z}^2 + \d x^2 + \d y^2
    \nonumber \\
    &\sim -\left( 1+\frac{Q^2}{\tilde{z}^2} \right) \d t^2 +\left(1+\frac{Q^2}{\tilde{z}^2} \right) \d\tilde{z}^2 + \d x^2 + \d y^2 .
\end{align}
The first arrow represents a coordinate transformation in the $\rho$-direction introducing the new coordinate $\tilde \rho$, through $\d \tilde{\rho}/ \d \rho=\left(1+Q^2/\rho^2\right)^{-1}$. For convenience, we express the result using $\rho (\tilde{\rho})$.  The second arrow means that we zoom in on $x= \rho \phi$ and $y= \rho \theta$ small, with the angles $\phi$ and $\theta$ small.  We can then express the metric using Cartesian coordinates where we write $\tilde{z}=\tilde{\rho}$. Similarly, it is natural to write $z=\rho$. Finally, since $Q^2/z^2$ is small, we can put $z \sim \tilde{z}$. Now let us see if we can match this to the induced metric on a brane embedded into (\ref{eq: constE backreact}).

Since the electric field giving rise to the metric in (\ref{eq: constE backreact}) is assumed to be constant (for simplicity), we can only hope to match the induced metric in (\ref{eq: RN}) over a short interval. We choose to study the embedding close to $\tilde{z}=\tilde{z}_1$, where $\tilde{z}_1$ is far away from the source such that $Q/\tilde{z}$ is small. This means that $r(\tilde{z})= r_1 + \alpha (\tilde{z}-\tilde{z}_1)$ is almost constant with $\alpha$ small. We set the asymptotic value of $r$ at infinity to $r_0$. Noting that the $\d r^2$-piece of the bulk metric only contributes at quadratic order in $\alpha$ to the induced metric, we see that the induced metric can be put in the form (\ref{eq: RN}) provided that
\begin{equation}
    \begin{split}
        k^2 r^2 -\frac{2\gamma^2 \log (k\zeta r)}{3k^{2}r^{2}} &= k^2 r_0^2\left( 1 + \frac{Q^2}{z^2} \right),\\
    k^{2}r^{2}+\frac{\gamma^{2}\log (k \zeta r)}{3 k^{2} r^{2}} &= k^2 r_0^2 .
    \end{split}
\end{equation}
The first equation comes from matching $\d t^2$ and $\d z^2$, while the second comes from matching $\d x^2$ and $\d y^2$. Here, we have already made a conformal rescaling of $x$, $y$ and $z$ to match the equations at zeroth order. This gives
\begin{equation}
    \begin{split}
        \frac{r^2}{r_0^2}&=1+\frac{Q^2}{3z^2}, \\
    \frac{\gamma^{2}\log (k \zeta r)}{3 k^{2}r^{2}} &= -\frac{k^2 r_0^2Q^2}{3z^2}.
    \end{split}
\end{equation}
Due to our approximation for the bulk metric, using a constant electric field, these equations are only valid close to a given point such as $z=z_1$. The first of the two equations give us $r$, for a given $z$, while the second equation determines $\zeta$.\footnote{Changing $z_1$, leads to a different $\zeta$ as a consequence of the approximation.} We immediately see from the first equation that $r$ grows in the direction towards the charge sourcing the field. The interpretation is simple. In order to capture the blue shift of the induced metric as we approach the source, the brane must bend upwards to make use of the bulk blue shift due to AdS. In this way the induced metric comes out right. We expect the result to be similar if we were to construct the full uplift of an electric field decaying away from a point source.

To better understand what is going on, let us recall the case of a point mass in 4D as discussed in \cite{Banerjee:2020wov}, caused by a string pulling on the brane. The brane bends upwards due to the pulling. The bending in itself causes a change in the vacuum solution of the bulk, with the induced metric on the brane becoming that of Schwarzschild. Here, we have a similar story even when ignoring the presence of any pulling string. The brane bends upwards in the direction of increasing electric field but now it is caused by the presence of the bulk B-field. This gives rise to a backreaction in the bulk that makes the brane bend upwards.

\section{A dark bubble consistency proof}\label{sec: check}

The dark bubble construction is based on two sets equations: Einstein's equations in the bulk (including the junction conditions) together with the Gauss-Codazzi equations. The 4D Einstein equation is obtained from the Gauss-Codazzi equation making use of the junction conditions. Schematically, the 4D Einstein tensor is:
\begin{equation}
    G_{4D} = T_{\rm extrinsic} - T_{\rm brane}\,.
\end{equation}
When setting up the equations for a particular physical situation, we have seen how certain choices need to be made. In particular, we have studied two examples involving electromagnetic fields in this paper. We had to choose initial conditions that were such that the 4D effective theory did not involve any additional fields beyond those that were represented on the brane itself. Note that in general such dark components can be present in the bulk, affecting the 4D physics. The dark radiation due to an AdS-Schwarzschild black hole, as already discussed in \cite{Banerjee:2018qey}, is one such example. The cosmological constant itself, can be viewed as another. 

What is crucial is that this initial condition is preserved by the time evolution. The central claim of the dark bubble proposal is, in fact, that this is the case. The initial condition that we impose is 
\begin{equation} \label{eq: Einstein is brane}
    G_{4D} = T_{\rm brane} ,
\end{equation}
implying that $T_{\rm extrinsic}= 2 T_{\rm brane}$ initially. The reason why this remains true for the subsequent time evolution is twofold:

\begin{enumerate}
    \item $T_{\rm brane}$ is covariantly conserved on its own through its equations of motion derived from the brane action. Therefore, there cannot be any exchange of energy between $T_{\rm brane}$ and anything else. 
    \item The total energy momentum tensor, involving $T_{\rm extrinsic}$ is also covariantly conserved since it is equal to the 4D Einstein tensor. Hence, the Einstein equation in the form (\ref{eq: Einstein is brane}) is preserved throughout.\footnote{One might worry, that there could be spontaneous creation of modes with $G_{4D}=0$, thus not visible to the Einstein equations and constrained by our argument. In writing down the Gauss-Codazzi equations, we have only employed them at the level of the Ricci-tensor when obtaining the 4D Einstein equations. However, they are valid at the level of the Riemann tensor, which then constrains such waves. This was fully explored in \cite{Danielsson:2022fhd}, where the uplift of 4D gravitational to 5D was worked out in detail.}
\end{enumerate}

Physically, this means the following. Let us assume matter associated with the brane such as the electromagnetic fields studied in this paper. This could also include other matter of the standard model represented by more elaborate brane configurations. We will call this the {\it visible} sector. Then set up initial conditions on the brane and in the bulk to make sure that the induced 4D energy momentum is precisely the expected one of the visible sector. Our discussion then shows that the subsequent time evolution will be the one of the visible sector coupled to 4D gravity. Any initial mismatch between the energy momentum tensor of the visible sector and the effective induced 4D energy momentum tensor, we interpret as the presence of a {\it dark} sector. This implies that this dark sector will, from the 4D point of view, couple to 4D gravity in the same way as the visible sector. It is natural to associate the cosmological constant with the dark sector, but it is clear from the rich structure of the model that other more dynamical components associated with the extra dimensions can be turned on as well. It would be interesting to explore whether these could be candidates for dark matter.

Note that there are corrections to the 4D Einstein equations at very high energies. As discussed in \cite{Banerjee:2019fzz}, they arise through quadratic couplings of the brane energy momentum tensor to itself.\footnote{A simple way to see this is through footnote 1 of \cite{Danielsson:2021tyb}. Just add a contribution to the brane tension $\sigma$.} In terms of the Friedmann equation, the standard contribution is of the form $l_4^2 \rho$, where $\rho$ is the energy density. The corrections are of the form $l_4^4 L^2 \rho^2 \sim (l_4^2 L^2 \rho ) \times l_4^2 \rho$, as indicated in equation (\ref{eq: ProjectedEinsteinEqs}). Using the relation between the scales, we find that $l_4^2 L^2 \rho \sim l^4_{10} \rho$. Hence, these corrections only appear at densities close to the 10D Planck scale (or string scale).

To further understand how tightly constrained the construction is, recall how the matter equations of motion are built into Einstein's equations. To illustrate, let us consider an elementary example in the form of a charged, thin spherical shell in vacuum. The shell is uniformly charged and has some tension. Effects due to gravity are assumed to be so small that they can be ignored. The evolution of the shell is easily obtained using the laws of electromagnetism. If any radial velocity of the shell is small, it is enough to use electrostatics involving electric forces or electric potentials. There is, however, another method that can be used. View it as a problem in general relativity, introducing a Reissner-Nordstr\"om metric on the outside and flat space time on the inside, together with a junction condition. Solving these equations, and taking the limit where the gravitational effects go to zero, you recover the elementary results obtained from electrostatics. This demonstrates how the Einstein equations know of, and take fully into account, the equations of motion of any matter in the system. This was also the content of the calculation done in \cite{Danielsson:2022fhd}, where it was shown how the brane equations of motion written in their Hamiltonian form were nothing other than the junction conditions.
\section{Discussion and outlook}\label{sec: discussion and outlook}

In this paper we have taken the first step towards explicitly incorporating the standard model of particle physics into the dark bubble model by considering electromagnetism. It is interesting to see the crucial role played by the B-field. It yields the necessary back reaction on the brane in order to get the proper 4D gravity. In many ways, the extended structure of B-fields into the bulk, and their back reaction on the 5D space time, play a very similar role to the stretched strings required to sustain the dust particles mentioned in the beginning. In the setting we have discussed in this paper, the backreaction of the electromagnetic field on the brane is driven by the B-field, which controls the jump in eq \eqref{eq:Hjump}. However, in principle the junction conditions could be modified even without B-fields in the theory. This is what we expect in string constructions arising \emph{e.g.} from type 0 orientifolds. It would be interesting to investigate such alternative realizations of the dark bubble in future work. 

Our work represents an important first step toward building particle phenomenology within the dark bubble scenario. From a bottom-up perspective, the next step involves realizing non-abelian gauge fields, possibly including the Higgs mechanism. Given the results that we have presented, we expect a similar story if we consider a non-abelian extension. The bubble wall would then consist of several branes on top of each other, with Chan-Paton factors furnishing the appropriate gauge degrees of freedom for quarks and other fields. In case of the electroweak force, we would have two branes separared by a distance proportional to the Higgs field. The simplest realizations of this setting would not correspond to the precise Higgs mechanism that takes place in the standard model, since the scalar fields on brane stacks typically take values in rank-two tensor representations (the adjoint in supersymmetric cases). The most natural way to separate the branes would be to keep them on top of each other in AdS at the same coordinate $r$, but at different positions on the internal $S^5$. Note that the mass of a string stretching between two such branes would be given by $l \times 1/l_s^2$, where $l$ is the length of the string, with the maximal value equal to the the Planck mass when the branes sit at opposite poles. It would be very interesting to work out the details of the B and H-fields, and their dependence on the $S^5$. This should give rise to a qualitatively different behaviour in $AdS_5$, compatible with massive particles (the W and Z) from the 4D point of view.

It is tantalizing that the string scale (and 10D Planck scale) ends up at around $10$ TeV, which is just above what is currently accessible in experiments. This suggests that massive excitations of open strings attached to the brane, such as those giving rise to \emph{e.g,} photons, could be detectable in future collider experiments. As we have discussed, the dark bubble model also includes a possible candidate of dark matter through the its dark sector, when a mismatch with the visible sector is introduced through the extrinsic curvature of the brane embedding. We hope to return to these phenomenological aspects in the near future.

\section*{Acknowledgement}

We thank Souvik Banerjee for collaboration at an early stage of this work, and Thomas Van Riet for discussions. DP would like to thank the Centre for Interdisciplinary Mathematics (CIM) for financial support. The work of SG was conducted with funding awarded by the Swedish Research Council grant VR 2022-06157.

\small
\bibliography{biblio}
\bibliographystyle{utphysmodb}
\end{document}